\documentclass[12pt]{styles/iopart}

\usepackage{graphicx} 
\usepackage{subcaption} 
\usepackage[numbers]{natbib}
\usepackage{siunitx}
\usepackage{float}
\usepackage{hyperref}

\makeatletter
\def\equation{\@beginparpenalty\predisplaypenalty
 \@endparpenalty\postdisplaypenalty
\refstepcounter{equation}\trivlist \item[]\leavevmode
 \hbox to\linewidth\bgroup $ \displaystyle
\hfil}
\def\endequation{$\hfil \displaywidth\linewidth\@eqnnum\egroup \endtrivlist}
\makeatother

\providecommand{\newblock}{}

\begin{document}

\title[]{A facility for thermo-mechanical characterization of fusion magnet materials during cryogenic ion irradiation}

\author{A Aurora$^1$, AR Devitre$^{1,2}$, APC Wylie$^{2}$ JA Rajagopal$^{2}$, MP Short$^{2}$}
\address{$^{1}$ Plasma Science and Fusion Center, Massachusetts Institute of Technology, Cambridge, 02139 MA, USA}
\address{$^{2}$Department of Nuclear Science and Engineering, Massachusetts Institute of Technology, Cambridge, 02139 MA, USA}

\ead{aaurora@mit.edu}
\vspace{10pt}
\begin{indented}
\item[]December 2024
\end{indented}

%
%
%
%
%

\begin{abstract}

Commercial fusion power plants demand magnet materials that retain structural integrity and thermal conductivity while operating under the bombardment of energetic neutrons at cryogenic temperatures. Understanding how thermo-mechanical properties evolve under such extreme loads is crucial for selecting materials with high radiation tolerance and predictable failure mechanisms. Presented here is a facility that combines cryogenic transient grating spectroscopy (TGS) with simultaneous ion irradiation, enabling \textit{in-situ} measurements of thermal diffusivity and surface acoustic wave (SAW) frequency spectra. Employing copper as a benchmark material, an irradiation was performed at 30\,K with 12.4\,MeV $\text{Cu}^{6+}$ ions producing a final fluence of $1.9 \times 10^{17}$\,ions/m$^2$. Over the irradiation period, thermal diffusivity nearly halved from an initial value of $1.2 \times 10^{-4}$\,$\text{m}^2/\text{s}$ while SAW speed did not show significant change, maintaining a value of $2162\pm18$ \,m/s. Given its real-time monitoring capability and the numerous candidate materials that remain uncharacterized under fusion magnet operating conditions, this facility is poised to deliver new scientific insights into fusion magnet material degradation trends, contributing to improved design criteria and operational certainty for forthcoming fusion power plants.

\end{abstract}
\section{Introduction}
\label{sec:introduction}

At present, progress toward commercial fusion energy is limited by the radiation tolerance of solid components surrounding the core. Neutrons born in fusion reactions cause atomic displacements and transmutations in the crystal structure of the constituent materials, degrading their properties, until the components must be replaced. Of particular interest to fusion power plant (FPP) economic viability is the lifetime of superconducting magnets, responsible for the magnetic field that sustains fusion reaction in the core. Due to their prohibitive cost, fusion magnets are typically considered lifetime components, i.e., intended to be non-replaceable. Over time, neutron bombardment degrades these magnets' ability to generate the confinement field efficiently, setting the operational lifetime of the FPP itself and its ultimate return on investment \cite{Nicholls2022, Dose2021}. 

The relevant literature has primarily focused on the electrical properties of magnet materials \cite{Devitre2024, Iliffe2021, Fischer2018}, leaving a notable gap in the understanding of thermo-mechanical property evolution under FPP operating conditions. Specifically, current research on the effects of irradiation at cryogenic temperatures remains insufficient for (1) selecting fusion magnet materials that ensure power plant safety and reliability under large cyclic and transient stresses typical of a fusion power plant, (2) assessing these materials' performance degradation, (3) making data-driven engineering design decisions, and (4) deriving high-level techno-economic metrics to maximize the commercial viability of fusion \cite{Knaster2016}. To mitigate these uncertainties, the fusion community must deliberately accelerate and scale material characterization studies in coupled conditions.

Transient grating spectroscopy (TGS) has emerged as a promising method for this purpose. TGS is an optical technique that provides rapid diagnoses of thermo-mechanical properties, is inherently non-destructive, and operates on the length scales of ion irradiation damage \cite{Dennett2019, Hofmann2019}. Unlike traditional \textit{ex situ} approaches—such as nanoindentation, X-ray diffraction, or TEM—that cannot easily operate in extreme irradiation environments, TGS enables \textit{in situ} measurements. And amongst techniques such as \textit{in situ} TEM \cite{Hattar2014} or \textit{in situ} Raman spectroscopy \cite{Miro2016} that can provide local structural information during irradiation, TGS is unique in its ability to continuously monitor thermal and elastic properties.

In practice, TGS functions by measuring diffracted light from an optically-induced thermal grating induced on a material surface. As the grating evolves, changes in the diffracted light pattern are used to infer elastic and thermal transport material properties.

Previously, TGS has been employed in tandem with ion-beam accelerators that emulate neutron irradiation \cite{Was2014, Hofmann2015}.  Short and Dennett et al. \cite{Short2015, Dennett2019} introduced \textit{in situ} ion irradiation with TGS to infer bulk property changes as they occur, enabling the quantification of relationships between dose and void swelling \cite{DennettVoidSwelling2018}, elastic moduli \cite{Dennett2016}, and thermal diffusivity \cite{Wylie2022} in single-crystal metals. Separately, prior art on cryogenic TGS investigated thermal \cite{Pennington1993}, spin \cite{Mahmood2018}, and acoustic \cite{Kim2021, Huberman2019} transport property evolution in several solid-state materials \cite{Choudhry2021}. 

However, to our knowledge, TGS has never been used at cryogenic temperatures with simultaneous irradiation. Developing this capability is important for three main reasons. First, it enables the coupled analysis of material evolution under fusion magnet operating conditions, specifically temperatures less than 30 K \cite{Mumgaard2017} and irradiation doses greater than 4 millidisplacements per atom (mdpa). This threshold is especially relevant because it is the expected cumulative radiation dose that rare-earth barium copper oxide (REBCO)—the primary constituent of high-temperature superconducting (HTS)—can withstand over their operational lifetime in fusion magnets \cite{Iliffe2021}. Secondly, the temperature at which irradiation occurs influences the resulting defect structures formed in the material. If an irradiation occurs above the migration temperature of a defect, then defects can move about during the irradiation and possibly annihilate or agglomerate. In most cases, this defect state results in property changes that differ from those caused by low-temperature irradiation \cite{Brown1981}. Therefore, only irradiations performed near the operating temperature of the magnet are unambiguous and, thus, useful for predicting changes in magnet material properties \cite{Devitre2024}. Finally, simultaneous measurement enables high-resolution mapping of temperature-dose-property relationships, at a fidelity and speed previously unattainable with more traditional \textit{ex situ} methods. This capability allows TGS the observation, either directly or by inference, of emergent microstructural evolution transitions as they occur, such as those during defect annealing.

This work introduces an apparatus constructed specifically for variable-temperature TGS, ranging between 30 K and 300 K. To demonstrate this capability and understand the impact of different heat loads (conductive, convective, TGS laser, and ion beam) on TGS response, copper samples are analyzed over the full temperature range without ion irradiation and a heat load study is conducted (\ref{appendix:head_load_study}). An irradiation is then performed at 30~K, showcasing the thermal diffusivity and sound speed evolution of Cu-110 under self-ion irradiation.

Cu-110 was selected as a benchmark material for its crucial role in  REBCO HTS magnets and cables \cite{Hartwig2020, Hartwig2024}. For example, VIPER cables (operating at 20~K) exhibit high cryostability, benefiting from copper specific heats and thermal conductivities. Within the constituent REBCO tapes, the copper stabilizing layer also serves as a quench protection mechanism by dissipating heat and providing an alternative current path during transient events \cite{Fu2003}. But these notions of copper cryo-stability are not guaranteed under the effects of irradiation, and copper alloys have demonstrated severe thermal diffusivity degradation and embrittlement post-cryogenic irradiation \cite{Fabritsiev1996}. To better understand the nature of such material degradation and validate the feasibility of cryogenic TGS, a bespoke apparatus and experiment are detailed in the forthcoming sections.

\section{Method}
\label{sec:methods}

\subsection{Transient grating spectroscopy (TGS}
\label{sec:tgs}

TGS is a non-contact method for characterizing changes in elastic, acoustic, and thermal transport properties, making it a powerful tool for inferring material structure evolution \cite{Hofmann2019}. Its sensitivity to micron-scale features makes TGS particularly well-suited for examining material degradation under ion irradiation, where damage typically manifests at comparable spatial scales.

\begin{figure}[htbp]
    \centering
    \begin{subfigure}[b]{0.4\linewidth}
        \includegraphics[width=\textwidth]{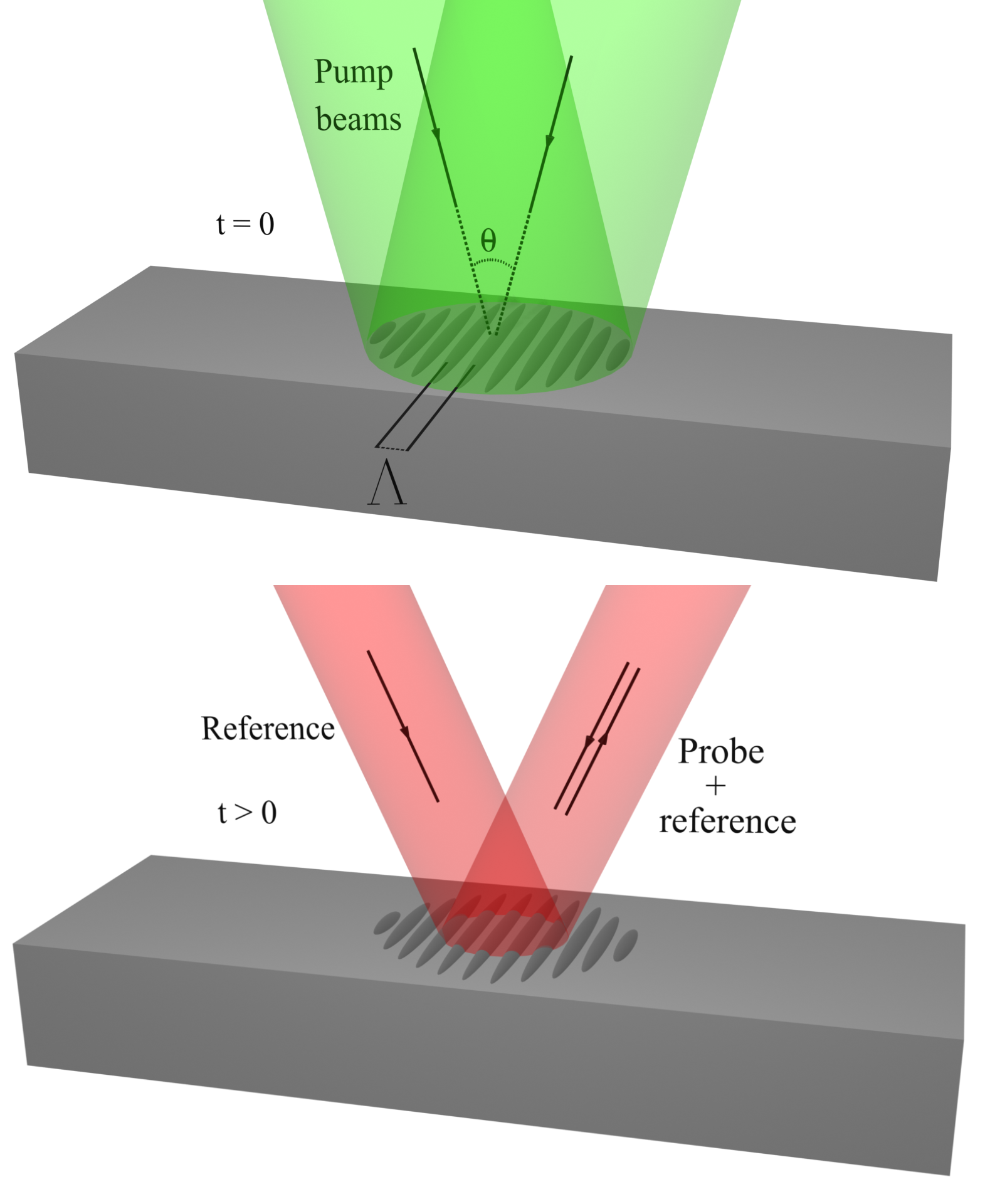}
        \caption{Transient grating formation and observation}
        \label{fig:tgs_fundamentals}
    \end{subfigure}
    \begin{subfigure}[b]{0.5\linewidth}
        \includegraphics[width=\textwidth]{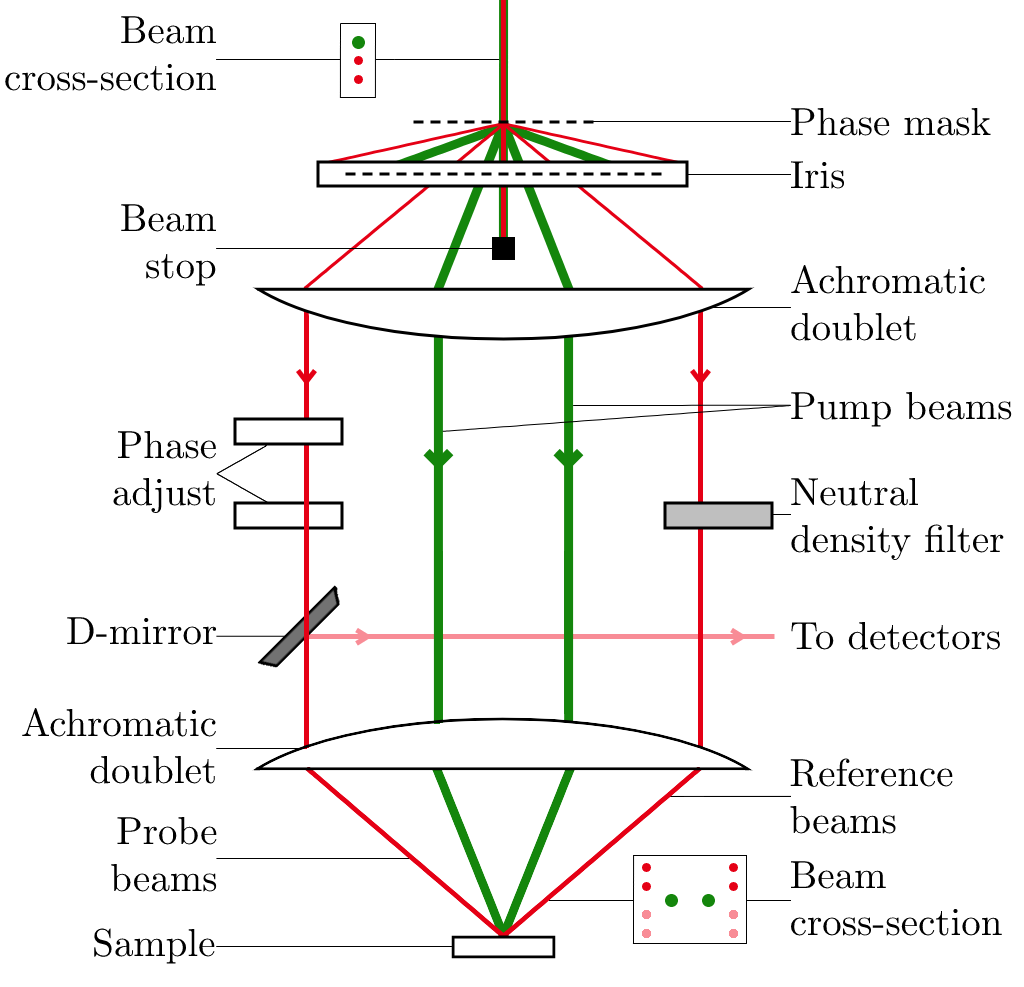}
        \caption{Di-homodyne TGS schematic}
        \label{fig:tgs_schematic}
    \end{subfigure}
    \caption{Schematic illustrating the di-homodyne TGS setup. At $t = 0$, a transient grating forms on the sample surface by the interference of two pump beams. At $t > 0$, the grating decay is observed by a reference and probe beam. Figure reproduced from Wylie et al. \cite{Wylie2023}.}
    \label{fig:tgs}
\end{figure}

TGS initiates with two short excitation laser beams coinciding on a polished sample surface creating a periodic intensity pattern known as the transient grating (Figure\,\ref{fig:tgs_fundamentals}). Constructive interference and light absorption within the beam overlap cause local heating in the same periodic pattern, forming a thermal grating. This heating gives rise to rapid physical displacement due to thermal expansion and a change in surface reflectivity. This periodic displacement also generates counter-propagating surface acoustic waves (SAWs). The decay of the thermal grating and SAW propagation is continuously monitored using a probe laser beam diffracted from the transient grating \cite{Kading1995}. In this experiment, the intensity of the diffracted probe beam is spatially overlapped with a reference beam reflected from the sample surface and used in a di-homodyne amplification scheme (Figure\,\ref{fig:tgs_schematic}) \cite{Maznev1998}. 

In TGS theory, the inducted surface displacement field $u_z(x,t)$ follows the form of an error function over time \cite{Kading1995}:
\begin{equation}
u_z(x,t) \propto \mathrm{erfc} \left( q \sqrt{\alpha t} \right)
\label{eq:surface_displacement}
\end{equation}
while the surface thermal field $T(x, t)$ decays exponentially:
\begin{equation}
T(x, t) \propto \frac{1}{\sqrt{t}} e^{-q^2 \alpha t}
\label{eq:thermal}
\end{equation}
where $q = \frac{2\pi}{\Lambda}$ is the excitation wave vector (rad/m) with $\Lambda$ being the transient grating wavelength or spacing (m), and $\alpha$ being the thermal diffusivity (m$^{2}$/s) in the grating vector direction. In practice, $\Lambda$ can be tuned for variable depth of investigation by adjusting the angle of the pump beams relative to the sample normal, $\theta$, with the following relationship: 
\begin{equation}
L = \frac{\Lambda}{\pi} = \frac{\lambda}{2\pi \sin\frac{\theta}{2}}
\label{eq:tgs_depth}
\end{equation} 
where $\lambda$ is the optical wavelength (m) and $L$ is the TGS thermal investigation depth (m).

In order to account for the oscillations of SAW decay, Hofmann et al. \cite{Hofmann2015} also include a damped sinusoidal term, which has been shown to reduce fitting sensitivity. Together, these phenomena underpin the theoretical model for the TGS response given by $I(t)$:
\begin{equation}
I(t) = A \left[ \text{erfc}\left(q\sqrt{\alpha t}\right) - \frac{\beta}{\sqrt{t}} e^{-q^2\alpha t}\right] + B\sin(2\pi ft + \Theta)e^{-\frac{t}{\tau}} + C
\label{eq:tgs_response}
\end{equation}
where $A$, $B$, and $C$ are fitting constants (W/m$^{2}$, $\beta$ represents the ratio of contributions to the signal from surface displacement and reflectivity (s$^{1/2}$), $f$ is the SAW frequency (Hz), $\Theta$ is the acoustic phase (rad), and $\tau$ is the acoustic decay constant (s).

\begin{figure}[htbp]
\centering
\includegraphics[width=1.0\linewidth]{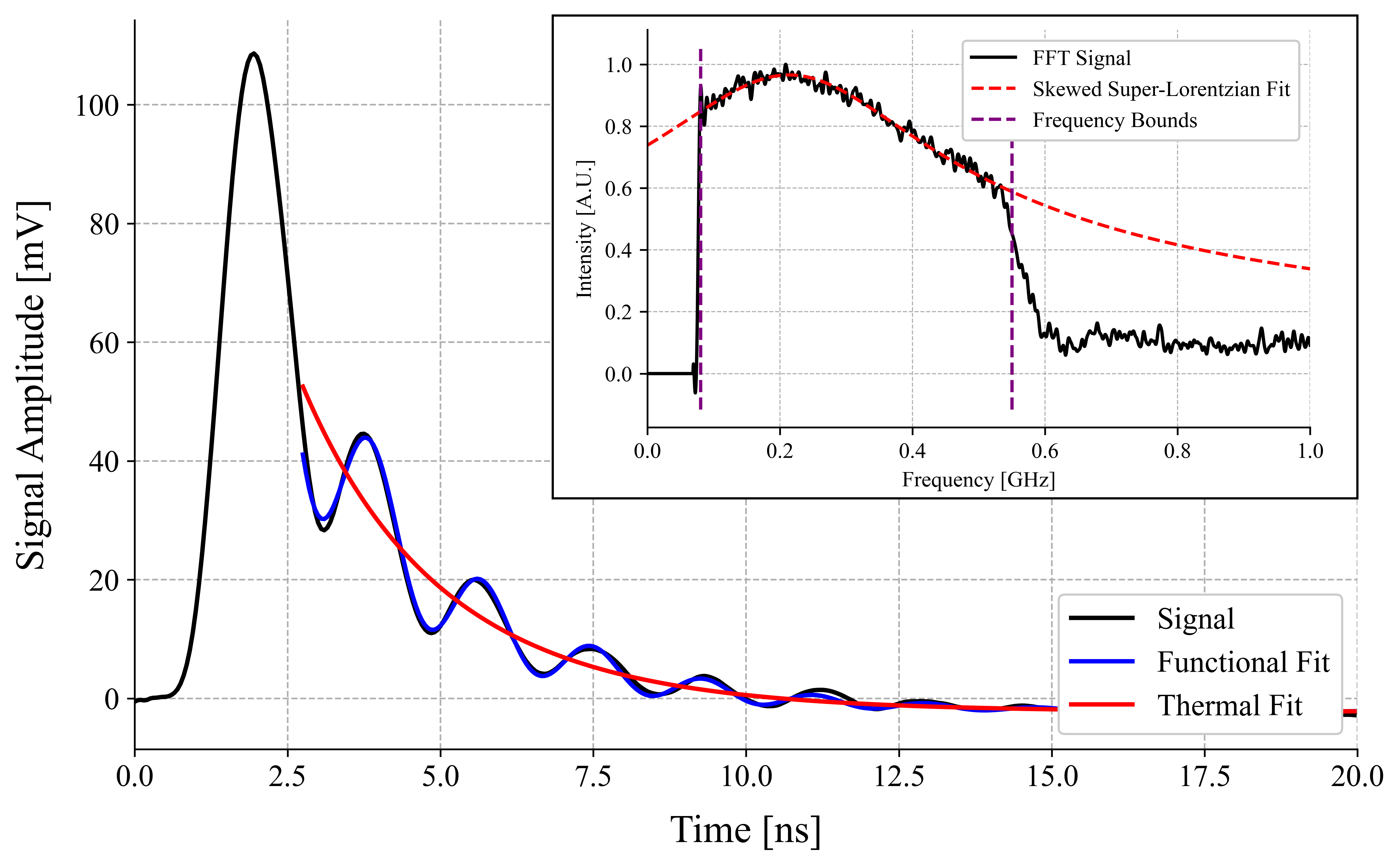}
\caption{TGS response for a polycrystalline Cu-110 sample at $20^{\circ}$\,C with a nominal grating spacing of 3.4\,$\mu$m. The full functional fit of Eq.\,\ref{eq:tgs_response} is shown in blue and the thermally governed grating decay is described by the red line, representing Eq.\,\ref{eq:tgs_response} without the acoustic term. The true signal response is collected over 200\,ns but truncated to 20\,ns for clarity. The inset displays the normalized Fast Fourier Transform (FFT) of the time-domain signal (black), along with a skewed super-Lorentzian fit (red dashed line) used to obtain initial estimates of the surface acoustic wave (SAW) frequency, $f$, and the acoustic decay constant, $\tau$. The purple dashed lines indicate the frequency bounds used for the analysis at 0.1 and 0.4 GHz.}
\label{fig:cu_trace}
\end{figure}

Figure\,\ref{fig:cu_trace} shows a typical TGS response for a polycrystalline copper sample. The signal comprises two distinct components: (1) an underlying decay attributed to the thermal equilibration of the inducted transient gratings in the sample and (2) periodic oscillations resulting from counter-propagating SAWs. Since $\alpha$ and $f$ are treated as parameters, fitting experimentally measured SAW decays directly affords the thermal diffusivity and SAW frequency, respectively. To obtain a measure of SAWs that is independent of SAW wavelength, SAW speed is the conventionally preferred metric. The SAW speed (m/s) can then be calculated using the formula $v = f \Lambda$. Repeating these measurements through time enables the data-rich characterization of a sample material's dynamically evolving properties \textit{in operando}.

The employed TGS apparatus uses a 532\,nm wavelength pump laser with a repetition rate of 1\,kHz, along with a 577\,nm quasi-continuous beam also made to pulse at 1~kHz using a chopper wheel and beam splitter to form two probe beams. Pump and probe beams were passed through a nominal grating spacing of 3.4\,$\mu$m, and the first-order pulses were imaged on the sample surface. Calibration using a \{100\}-oriented single crystal tungsten reference specimen (SAW speed: 2669.5\,m/s) revealed a true grating spacing of 3.5276\,$\mu$m~\cite{Dennett2016}. Probe beams are collected in two Si-avalanche photodiodes with a bandwidth between 50\,kHz – 1\,GHz (3\,dB) and are digitized with an oscilloscope. The final signal used for fitting was averaged over 10,000 pulses, and a no-pump baseline reading was subtracted to minimize inherent system noise.

TGS signal thermal fits were approximated using a Levenberg–Marquardt nonlinear least-squares fit after an initial na\"{i}ve fit used for parameter space restriction following the procedure described in~\cite{Dennett2018}. SAW frequency was estimated via a skewed super-Lorentzian fit to a fast Fourier transform of the TGS signal as shown in Figure\,\ref{fig:cu_trace}. The extracted parameters were subsequently used as initial guesses in a high-iteration round of Levenberg–Marquardt fitting, which refined the parameter estimates to produce the final results. Further rationale for this multi-step fitting procedure is discussed in Section~\ref{sec:results_thermal}. Scripts for fitting and raw TGS data are provided in Section~\ref{sec:availability}.

\subsection{Cryogenic target holder}
\label{sec:diagnostic_apparatus}

\begin{figure}[htbp]
    \centering
    \begin{subfigure}[t]{0.7\linewidth}
        \centering 
        \includegraphics[width=\textwidth]{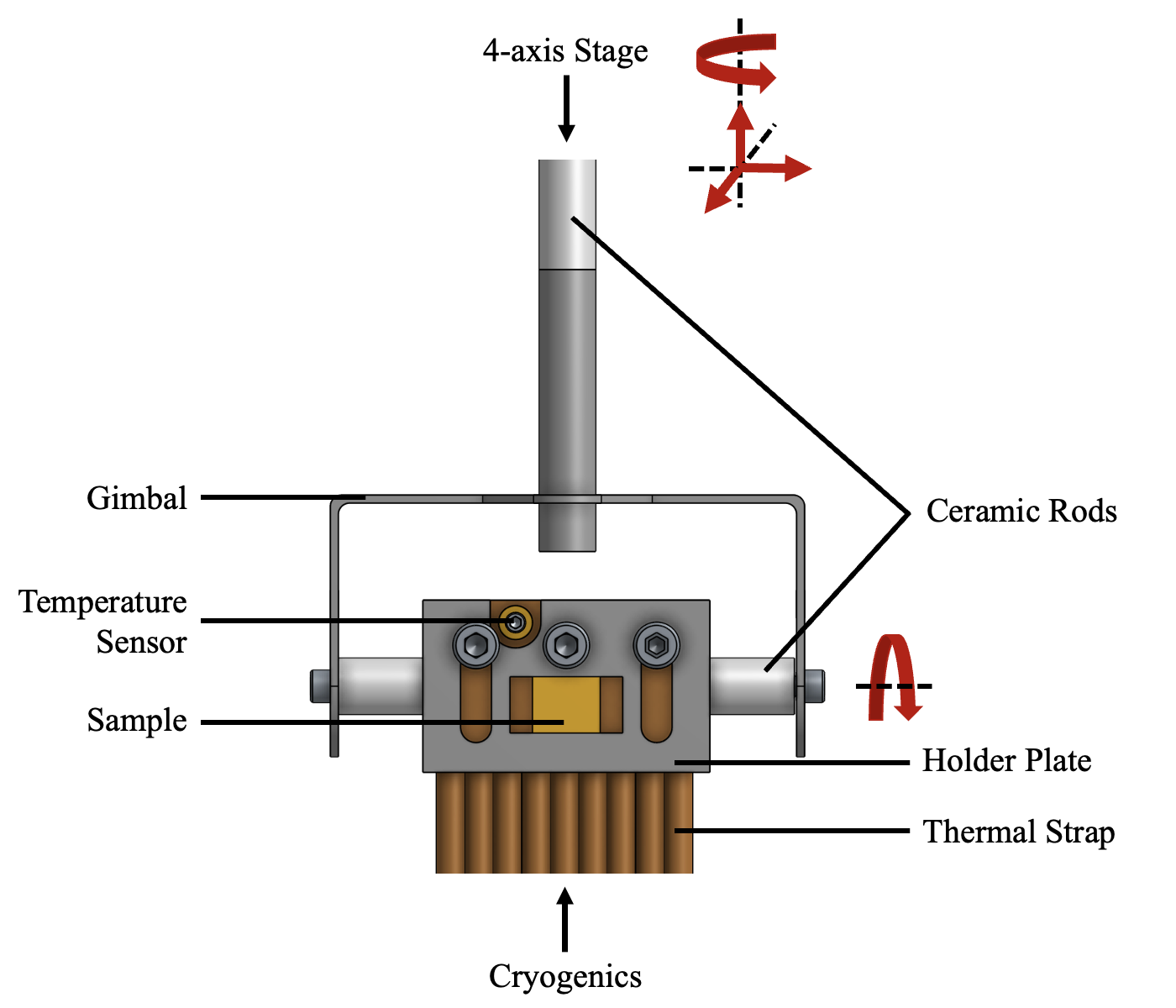}
        \caption{Target holder} 
        \label{fig:target_holder}
    \end{subfigure}
    \begin{subfigure}[b]{0.7\linewidth}
        \centering 
        \includegraphics[width=\textwidth]{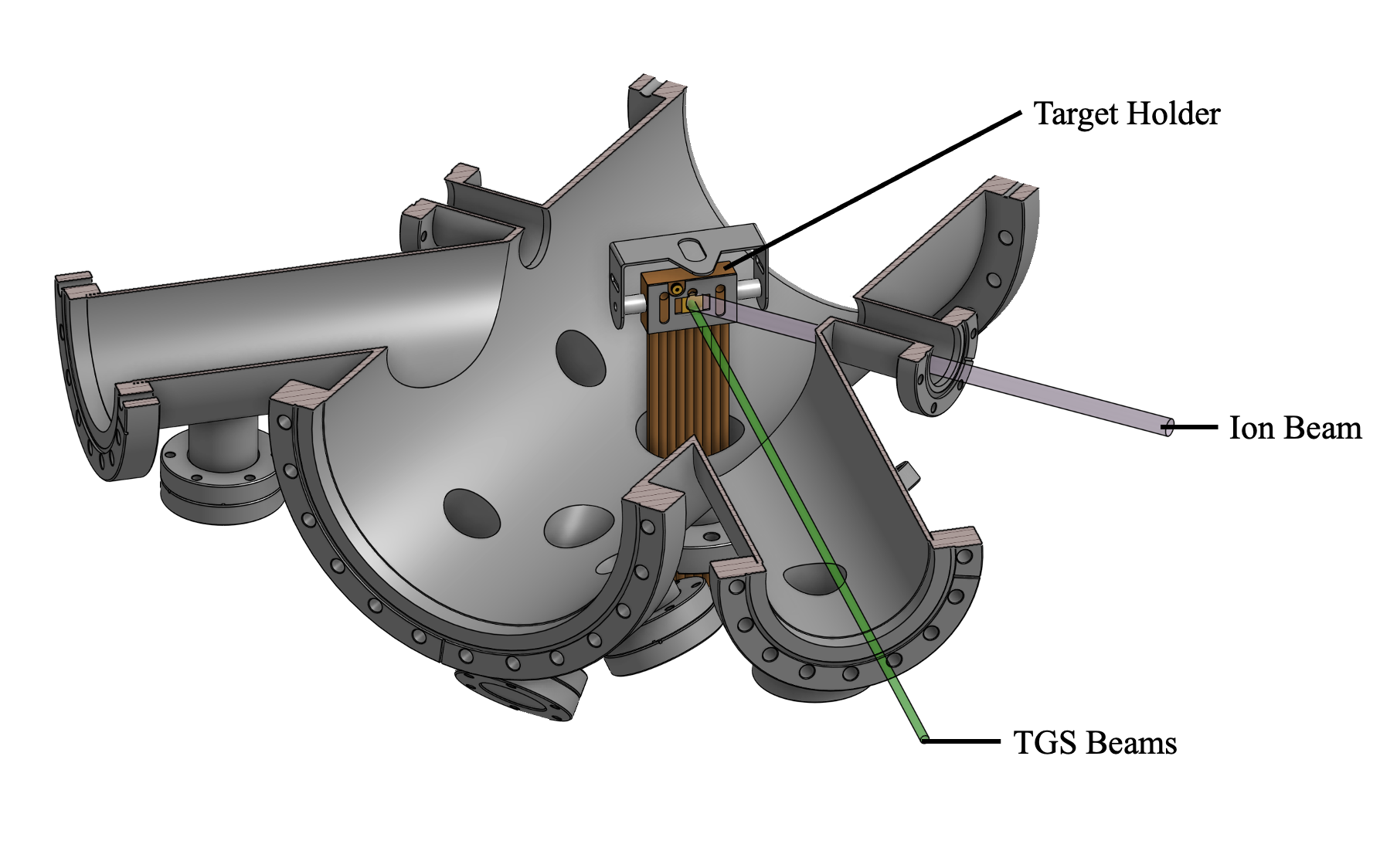}
        \caption{Vacuum chamber}
        \label{fig:cross-section}
    \end{subfigure}
    \caption{Experimental setup within the chamber: (a) A close examination of the target holder and its components. Here, ceramic rods are secured to the holder using set screws and are connected to the gimbal via a nut and two washers, facilitating pitch rotation (middle red arrow). From above, the holder is mounted to a 4-axis stage enabling XYZ translation and yaw rotation (top red arrows). Below, the holder, which dually functions as a flexible thermal strap, is coupled to input cryogenics for efficient cooling of the sample. Temperature measurements are recorded every second by a CERNOX sensor screwed into the target holder at a distance of 2\,mm from the sample edge. (b) A cross-section view of the target holder within the vacuum chamber. The TGS beams (green) and the ion beam (purple) are overlapped on the sample for simultaneous material characterization under irradiation.}
\end{figure}

The experimental apparatus combines a target holder (Figure\,\ref{fig:target_holder}) with an integrated system of cryogenics, TGS optics, and an ion-beam accelerator  (Figure\,\ref{fig:cross-section}). The interface offers 5-axis mobility for aligning the target perpendicular to the TGS pump beams while ensuring accessibility to the accelerator ion beam. The target holder is suspended from a 4-axis stage, exterior to the chamber, that enables motion along three spatial axes and yaw rotation. Inside the chamber, a movable steel gimbal, attached to the target holder with two glass-mica ceramic rods, facilitates pitch rotation. 

The sample is pressed against the flexible P50-502 copper thermal strap (Technology Applications Inc.) using an adjustable steel plate. The thermal strap serves as the target holder and interfaces with CVi’s CGR409 Cryocooler providing 5\,W of cooling power at a base temperature of 17\,K. A pressed indium contact is made at the strap/cryocooler interface to maximize thermal conductance. During experiments, the chamber is kept under vacuum, in the $10^{-4}$\,Pa range to minimize convection. The system achieves a base temperature of 30\,K with a cooldown time of 3 hours from room temperature. The local heating effects of the TGS and ion beams are also considered to determine whether these loads significantly elevated the sample temperature. Analytical calculations and experimental cooldown results (detailed in \ref{appendix:head_load_study}) confirm that the TGS beam's contribution to bulk sample temperature is negligible compared to the dominant radiative heat load. Localized ion beam heating effects, on the other hand, can be consequential; as is shown in a recent publication \cite{Devitre2025}, the temperature rise in the beam spot is not fully captured by nearby temperature probes and requires alternative means of quantification. However, a precise accountancy of ion beam point heating is beyond the scope of this work.

All experiments reported in this work utilized a 13\,mm x 15\,mm x 2\,mm Cu-110 sample (UNS C11000; 99.9\% Cu, 0-0.04\% O, 0-0.005\% Pb, 0-0.005\% Bi). The polycrystalline samples were cut from a cold-worked copper sheet acquired from McMaster Carr (PN: 9821K11). Each sample underwent a sequential polishing process, beginning with 800 grit sandpaper, followed by 1200 grit, 3\,µm diamond paste, 1\,µm diamond paste, and concluding with 40\,nm alumina oxide polishing suspension.

\subsection{Cu-110 irradiation case study}
\label{sec:irradiation_methods}

The MIT CLASS accelerator, a 1.7MV Tandetron with a source of negative ions by Cesium sputtering (SNICS), was utilized for the irradiation \cite{Devitre2024}. The beamline contains an integrated Faraday cup to measure the beam current. Upstream from the accelerator beamline, a beam profile monitor and quadrupole magnet system are used to maintain a Gaussian profile broadened over a 2~mm aperture to produce a spatially uniform beam spot on the sample. Alignment of the ion beam and TGS was performed by raising the stage until the TGS spot hit the sample clamp, which had been coated with ZnS:Ag scintillator. The ion beam was then made coincident with the TGS spot, with a horizontal offset made to allow for the vertical height of the clamp.

To examine the impact of cryogenic irradiation on the thermal diffusivity and SAW speed of Cu-110, the sample was subjected to 12.4\,MeV $\text{Cu}^{6+}$ ions with an average beam current of 0.45\,nA. Irradiation commenced alongside TGS at 30\,K for 30 minutes, resulting in a total fluence of $1.9 \times 10^{17}$\,ions/m$^2$, equivalent to 21 mdpa (as detailed in \ref{appendix:irradiation_damage_calculation}). Self-ion irradiation was chosen to avoid chemical contamination from extraneous ion implantation.

\begin{figure}[htbp]
\centering
\includegraphics[width=0.9\linewidth]{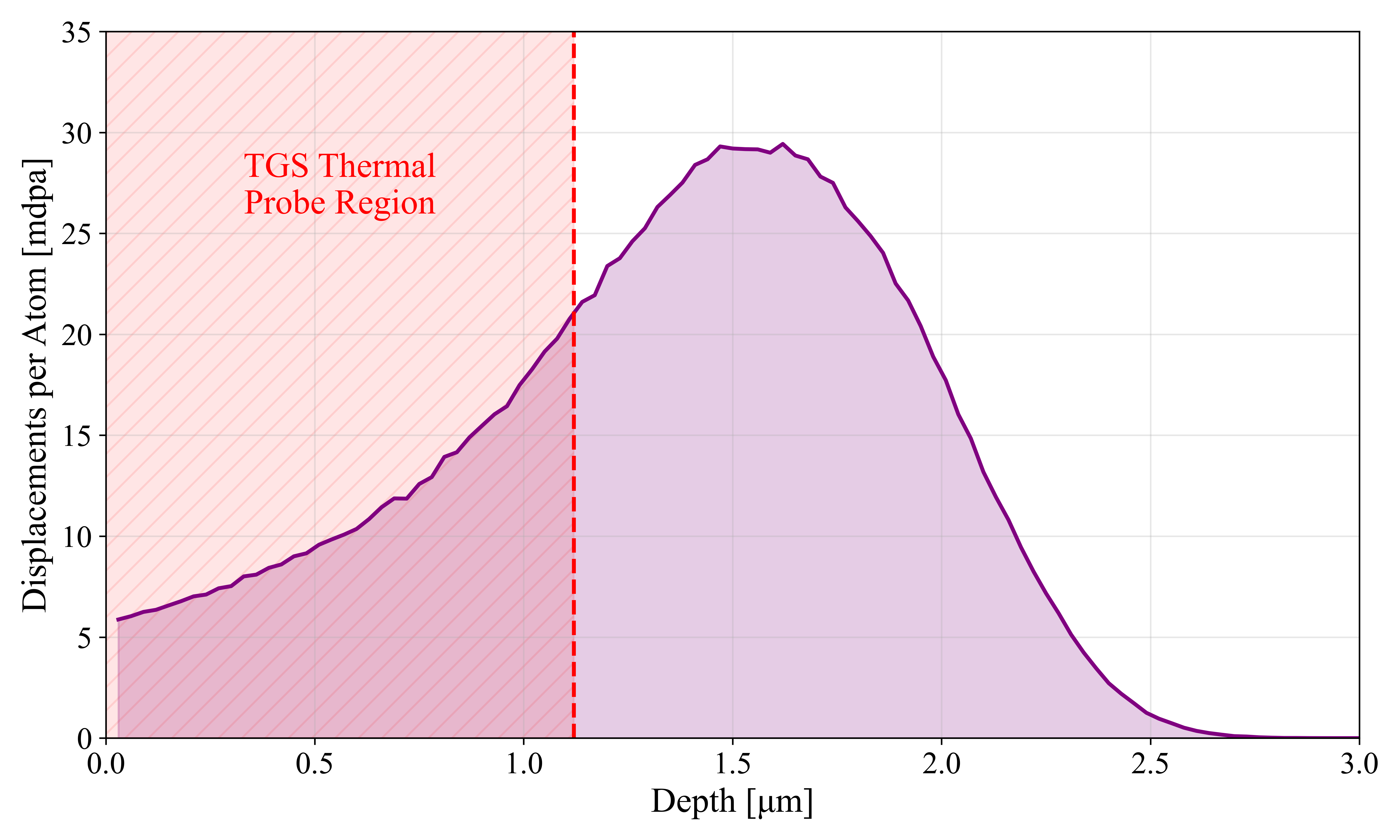}
\caption{SRIM simulated damage profile for 12.4\,MeV $\text{Cu}^{6+}$ into Cu-110 ($\rho = 8912.93$~kg/m$^3$) incorporating the $45^\circ$ ion incidence angle shown in Fig.\,\ref{fig:cross-section}. The y-axis shows binned displacements per atom (dpa), computed from the SRIM output (vacancies/Å-ion) using ion fluence and material density (see \ref{appendix:irradiation_damage_calculation}). The illustrated TGS thermal probe depth ($\Lambda/\pi$) reaches 1.12~$\mu$m where the mean displacement damage is 21~mdpa. While the theoretical probe region extends indefinitely, signal contributions decrease by a factor of $e$ at this depth.}
\label{fig:depth_dpa}
\end{figure}

Based on the aforementioned parameters and following ASTM-recommended settings, including a Cu displacement threshold energy of 30\,eV and the Quick Kinchin-Pease damage model, a SRIM simulation was conducted on a single 3\,\(\mu\)m layer of Cu with 100,000 ions \cite{Stoller2013, Ziegler2010}. Figure\,\ref{fig:depth_dpa} depicts the resulting displacement per atom (dpa) in the material as a function of depth.

\section{Results \& Discussion}
\label{sec:results_discussion}

\subsection{Thermal property evolution}
\label{sec:results_thermal}

\begin{figure}[htbp]
    \centering
    \includegraphics[width=\textwidth]
    {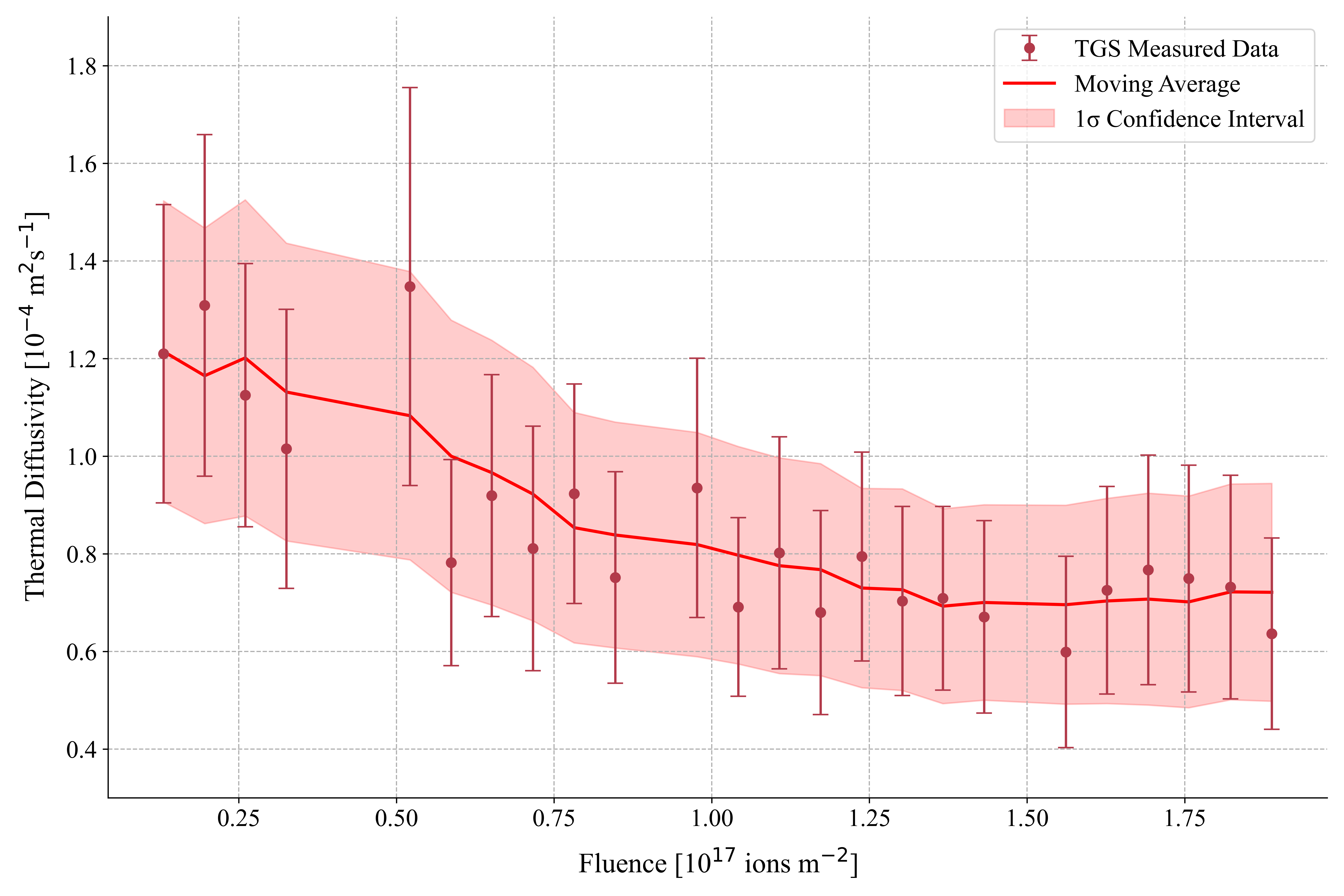}
    \caption{Thermal diffusivity of Cu-110 as a function of ion fluence measured by TGS at 30~K, showing a monotonic decrease with increasing irradiation. Error bars represent 1$\sigma$ statistical uncertainty, with the shaded region indicating the 6-point moving average and confidence interval. Outliers were removed using a chi-square test with a 95\% confidence interval.}
    \label{fig:alpha_irradiation}
\end{figure}

Consistent with previous high‐temperature irradiation studies on copper \cite{Trachanas2025} and other mono-atomic metals \cite{Wylie2022, Habainy2018}, thermal diffusivity decreases as radiation damage accumulates (Figure~\ref{fig:alpha_irradiation}). At 30\,K, the pre‐irradiation diffusivity is $1.2\times10^{-4}\,\mathrm{m}^2/\mathrm{s}$—about an order of magnitude below values reported for Cu with Residual Resisitivty Ratio (RRR)\,=\,30 ($\sim1\times10^{-3}\,\mathrm{m}^2/\mathrm{s}$) \cite{Duthil2014}. This discrepancy likely reflects how TGS's sensitivity is beneficial for detecting \emph{relative} changes in diffusivity, whereas its absolute calibration can be influenced by beam‐spot alignment, surface impurities, and other systematic effects.

As ion fluence increases, the measured diffusivity decays—nearly halving at $1.35\times10^{17}\,\mathrm{ions}/\mathrm{m}^2$ and plateauing afterwards at $0.7\pm0.22\times10^{-4}\,\mathrm{m}^2/\mathrm{s}$. The observed decrease in thermal diffusivity is possibly caused by stacking fault tetrahedra (SFT) defects, as suggested by previous room temperature, low dose irradiations \cite{Was2017, Li2013}. In metals, point defects act as scattering centers for free electrons, leading to a reduction in electrical conductivity, $\sigma$, as the density of point defects increases within the crystalline microstructure \cite{Trachanas2023}. Since electrical conductivity and thermal conductivity, $\kappa$, are interrelated by the Wiedemann-Franz Law, which holds for defected copper \cite{Ye2020}, a reduction in electrical conductivity is directly associated with a decrease in thermal diffusivity. This raises the possibility of using TGS as a non-contact method for resistivity recovery experiments, which has been one of the workhorses of radiation material science \cite{GomezFerrer2016}.

Error bars and global trend fluctuations in the data indicate thermal fitting uncertainty in the TGS signals. This uncertainty is exacerbated in copper samples due to its high thermal diffusivity amongst the metals, which results in fast decay of the thermal grating and a short signal lifetime. Consequently, the number of usable ``peaks" for thermal fitting is significantly reduced, an effect that is even more pronounced at cryogenic temperatures, where the TGS signal decays faster. To address this, a two-step fitting approach was developed: a) obtain an initial fit based on the technique outlined in Section\,\ref{sec:tgs} and then b) reapply the Levenberg-Marquardt algorithm using the initial fit to refine estimates for all parameters in Eq.\,\ref{eq:tgs_response}, including SAW frequency. Errors for the refined fits were recalculated using a covariance matrix of the fitting parameters and thus represent $1\sigma$. This approach improved both thermal and functional fitting sensitivity compared to existing techniques \cite{Wylie2022, Dennett2018}, especially in cases with noisy or rapid signal decay. 

Reasonable radiation operation limits for copper in Yttrium Barium Copper-Oxide (YBCO) HTS magnets are fast neutron fluences up to $2 \times 10^{21}$ $\text{n}/\text{m}^2$, corresponding to damage levels in the range of mdpa \cite{Zhai2018}. Concerningly, at ion fluences of just $0.6 \times 10^{17}$ $\text{ions}/\text{m}^2$ causing comparable order-of-magnitude displacement damage (6 mdpa), the copper sample's diffusivity at 30~K deteriorates by $17.9\pm6.75\%$. These initial findings indicate that more comprehensive property evolution studies are necessary to prove out existing assumptions regarding copper's cryostability and quench protection candidacy under cryogenic irradiation \cite{John2025}. 

\subsection{Elastic property evolution}
\label{sec:results_mechanical}

\begin{figure}
    \centering
    \includegraphics[width=\textwidth]
    {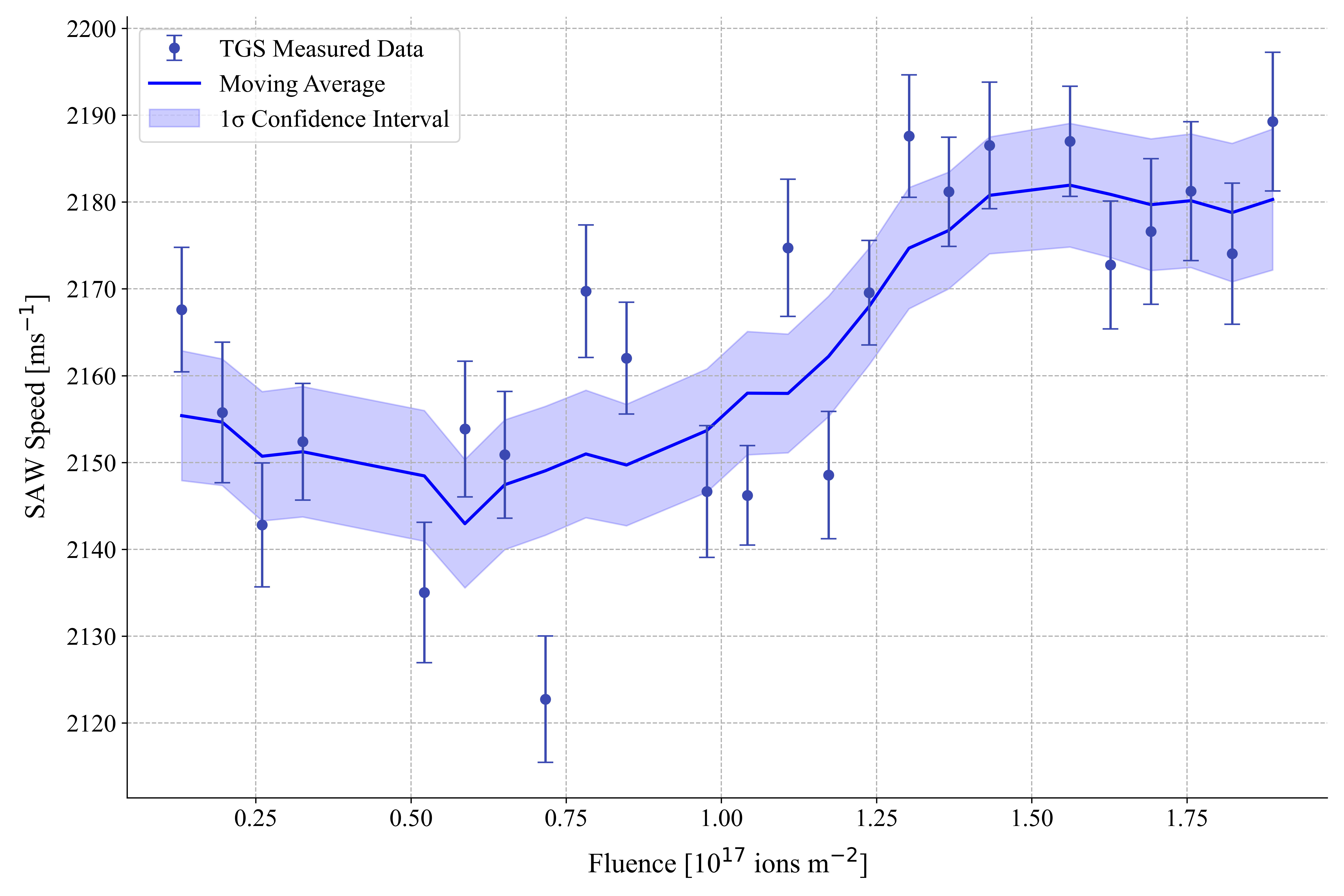}
    \caption{SAW speed of Cu-110 as a function of ion fluence measured by TGS at 30~K, showing little significant change. Error bars represent 1$\sigma$ statistical uncertainty, with the shaded region indicating the 6-point moving average and confidence interval. Outliers were removed using a chi-square test with a 95\% confidence interval.}
    \label{fig:SAW_irradiation}
\end{figure}

SAW speed on the TGS spot demonstrates minimal shift over the irradiation period. Initially, the SAW speed is $2155\pm7.2$\,m/s and remains relatively constant until reaching a fluence of $0.6 \times 10^{17}$\,ions/m$^2$. At this point, a more slight rise initiates, with the SAW speed reaching an average maximum of $2180\pm8.0$\,m/s at a fluence of $1.4 \times 10^{17}$\,ions/m$^2$ before restabilizing.

In metals and alloys like Cu-110, irradiation-induced lattice vacancies and interstitials form defect clusters within the microstructure. At cryogenic temperatures below 50 K, defect migration is minimized, meaning clusters are more stable and rarely recombine \cite{Zinkle2012, Wu2022}. A high cluster density impedes the motion and slip of dislocations within the crystal lattice. In metals, dislocation loops are the primary mechanism of plastic deformation, and their creation contributes to radiation hardening, correlating to an increase in material stiffness constants, a decrease in density, and an overall increase in SAW speed \cite{Shulman1970, Was2017}. However, the magnitude of these microstructural changes falls below the sensitivity threshold of the TGS probe, making them undetectable. 

While SAW speed trends offer a valuable relative understanding of irradiation-induced mechanical property evolution, the underlying single-crystal elastic constants are more useful for predicting this embrittlement process. Previous TGS studies on copper have utilized SAW speed measurements and theory grounded in the Christoffel equation to deduce stiffness constants in monocrystal samples subjected to irradiation \cite{Dennett2016, Dennett2018}. Recently, work by Du et al. \cite{Du2017} using electron backscatter diffraction (EBSD) mapping has also enabled the prediction of elastic constants in polycrystalline and anisotropic materials via TGS. Integrating this characterization approach with the presented \textit{in-situ} cryogenic irradiation method would enable direct measurement of coupled effects on the most fundamental mechanical properties.
\section{Conclusions}

We have built and tested a cryogenic target holder where the evolution of fusion magnet materials under irradiation can be assessed via non-contact measurements of thermal diffusivity and SAW speed. The \textit{in situ} diagnostic system enables the characterization of property evolution as it occurs, without intermediate annealing, in the temperature range of 30 to 300~K. Serving as a demonstration material, Cu-110 was irradiated with 12.4\,MeV $\text{Cu}^{6+}$ ions at 30\,K, up to a fluence of $1.9 \times 10^{17}$\,ions/m$^2$, corresponding to 21 \,mdpa. Over the irradiation period, SAW speed remained relatively constant while thermal diffusivity nearly halved, explained by the defect-induced scattering of conduction electrons. 

Measuring thermo-mechanical property changes during irradiation, at cryogenic temperature, is particularly important in the context of fusion, high-energy physics, and space technology where magnets will operate under the bombardment of high-energy particles. This apparatus aims to enable the collection of this data to inform predictive models, establish better design criteria, and facilitate practical engineering decisions. Future work with this device will focus on analyzing other material layers within REBCO superconductor tapes, monitoring temperature-controlled defect annealing in real-time, and potentially replacing resistivity recovery as the method of choice to investigate defect activation energies.
\section{Data availability statement}
\label{sec:availability}

The data and code that support the findings of this study are openly available in the following repository: \href{https://github.com/shortlab/2024-Cu-Cryo-TGS}{https://github.com/shortlab/2024-Cu-Cryo-TGS}. Generic Python scripts for TGS signal fitting are maintained and updated in the following repository: \href{https://github.com/shortlab/PyTGS}{https://github.com/shortlab/PyTGS}.

\section{Acknowledgements}

A Aurora would like to thank the accelerator gods for allowing beautiful beams during the final days of this project! And thank you to R Shulman and MP Short, for inventing FUSars and for all their invaluable guidance. All authors would like to thank Mr. Burns for sponsoring nuclear fusion research at MIT. 
\appendix

\section{Heat load study}
\label{appendix:head_load_study}

A critical requirement of the diagnostic apparatus is to reach the cryogenic range of interest while bearing experimental heat loads. To this end, the target holder was designed to maximize heat conduction and minimize heating from thermal radiation.

In addition, a heat load validation study was conducted to address uncertainty about how the TGS and ion beams would elevate the sample temperature. This involved an analytical calculation and validation experiment to decouple the contribution of the TGS and ion beams from other known heat loads.

\subsection{Analytical calculation}

The total heat load \(Q_{\text{H}}\) on the sample is estimated by considering the contributions from conduction, radiation, the TGS beam, and the ion beam. Thus, the formula for the total heat load is given by,
\begin{equation}
    Q_{\text{H}} = Q_{\text{cd}} + Q_{\text{rd}} + Q_{\text{cv}} + Q_{\text{tgs}} + Q_{\text{ion}}
\end{equation}

For \(Q_{\text{cd}}\), Fourier's law is applied using the room temperature thermal conductivity of 316LN stainless steel ($k_{\text{steel}} = 16.3\,\text{W/m} \cdot \text{K}$) and glass-mica ceramic ($k_{\text{ceramic}} = 1.46\,\text{W/m} \cdot \text{K}$), as cryogenic values were unavailable in the literature. The calculation considers different components' lengths and cross-sectional areas, such as the stage, gimbal, and rods as illustrated in Figure\,\ref{fig:target_holder}. The total thermal resistance, \( R_{\text{T}} \), is determined by combining the resistances of these elements in series and parallel, reflecting the cumulative path of heat conduction. The resultant heat load is calculated as,
\begin{equation}
    Q_{\text{cd}} = \frac{(T_{\text{A}} - T_{\text{S}})}{R_{\text{T}}}
\end{equation}

For \(Q_{\text{rd}}\), black-body radiation from the vacuum chamber interior walls is assumed to be driving and the Stefan–Boltzmann law is applied. The cylindrical vacuum chamber is approximated as a sphere with an equivalent diameter and the emissivity of 316LN stainless steel at room temperature ($\epsilon = 0.16$) is used,
\begin{equation}
    Q_{\text{rd}} = S\sigma\epsilon(T^4_{\text{A}} - T^4_{\text{S}})
\end{equation}

For \(Q_{\text{cv}}\), the load is assumed to be negligible since the system is in high vacuum on the order of $10^{-4}\,\text{Pa}$,
\begin{equation}
    Q_{\text{cv}} \approx 0\,\text{W}
\end{equation}

The heat loads of the TGS, \(Q_{\text{tgs}}\), and ion beam, \(Q_{\text{ion}}\), are characterized based on their respective experimental parameters, with the assumption of perfect absorption. Although this assumption is not empirically true, it provides a useful upper bound for the beams' contributions. For \(Q_{\text{tgs}}\), the heat input is directly equated to the laser power,
\begin{equation}
Q_{\text{tgs}} = P_{\text{tgs}} = 3.66\,\text{mW}
\end{equation}
For \(Q_{\text{ion}}\), the heat input is derived from the ion energy and beam current,
\begin{equation}
Q_{\text{ion}} = E_{\text{ion}}I_{\text{ion}} = 5.58\,\text{mW}
\end{equation}

The cooling capacity, \( Q_{\text{C}} \), of the cryocooler is a function of temperature as defined in the CVi Model CGR409 specification sheet. To retrieve the underlying parameters defining \( Q_{\text{C}} \), a plot digitizer is used to extract temperature and heat load coordinate pairs and fit a logistic function to the data. 

Setting \( Q_{\text{H}} \) and \( Q_{\text{C}} \) equal and inputting the relevant constant parameters, a function of the sample temperature, $T_{\text{S}}$, is obtained on both sides.
\begin{equation}
    Q_{\text{H}}(T_{\text{S}}) = Q_{\text{C}}(T_{\text{S}})
\end{equation}

Using the SciPy toolkit's Levenberg-Marquardt algorithm, this equation was solved numerically for several initial values and consistently produced, $T_{\text{S}} = 26.64\,\text{K}$. Plugging back into the heat load expressions, it was determined that $Q_{\text{cd}} = 0.354\, \text{W}$, $Q_{\text{rd}} = 8.656\,\text{W}$, and $Q_{H} = 9.02\,\text{W}$. Here, it becomes evident that the dominant heat load is radiative while contributions from conduction, the TGS beam, and the ion beam are comparatively negligible. 

Finally, accounting for the temperature drop across the thermal strap, which has conductance $C = 1.25  \, \text{W/K}$, 
\begin{equation}
    \Delta T = \frac{Q_{\text{H}}}{C} = 7.21 \, \text{K}
\end{equation}

We obtain the final sample temperature, $T_{\text{F}}$,
\begin{equation}
    T_{\text{F}} = T_{\text{S}} + \Delta T \approx 34 \, \text{K}
\end{equation}

\begin{table}[h!]
\centering
\renewcommand{\arraystretch}{1.2}
\setlength{\tabcolsep}{10pt}
\begin{tabular}{>{$}l<{$}p{0.55\textwidth}>{$}l<{$}}
\hline
\textbf{Symbol} & \textbf{Definition} & \textbf{Units} \\
\hline
Q_{\text{H}} & Total heat load & W \\
Q_{\text{cd}} & Conductive heat load & W \\
Q_{\text{rd}} & Radiative heat load & W \\
Q_{\text{cv}} & Convective heat load & W \\
Q_{\text{tgs}} & TGS beam heat load & W \\
Q_{\text{ion}} & Ion beam heat load & W \\
T_{\text{A}} & Ambient temperature & K \\
T_{\text{S}} & Sample temperature & K \\
R_{\text{T}} & Total thermal resistance & K/W \\
k & Thermal conductivity & W/(m\cdot K) \\
S & Surface area of vacuum chamber & m^2 \\
\sigma & Stefan-Boltzmann constant & W/(m^2\cdot K^4) \\
\epsilon & Emissivity & dimensionless \\
P_{\text{tgs}} & TGS beam power & W \\
E_{\text{ion}} & Ion beam energy (per unit charge) & V \\
I_{\text{ion}} & Ion beam current & A \\
Q_{\text{C}} & Cryocooler cooling capacity & W \\
\Delta T & Temperature drop & K \\
C & Conductance & W/K \\
T_{\text{F}} & Final temperature & K \\
\hline
\end{tabular}
\caption{Definitions and units of symbols used in the heat load study.}
\label{table:symbol_definitions}
\end{table}

\subsection{Validation experiment}

Experimentally, the cooldown curves of the sample with and without TGS were recorded. Temperature readings were sampled every second using a pre-calibrated LakeShore Cernox 1070 HT temperature sensor. The sensor was coupled to the target holder with a pressed screw and a thin layer of indium. After a 3-hour and 45-minute cooldown, the ion beam was introduced and measurements were recorded for another 40 minutes.

\begin{figure}[H]
\centering
    \begin{subfigure}[t]{0.95\linewidth}
        \centering  \includegraphics[width=\textwidth]
        {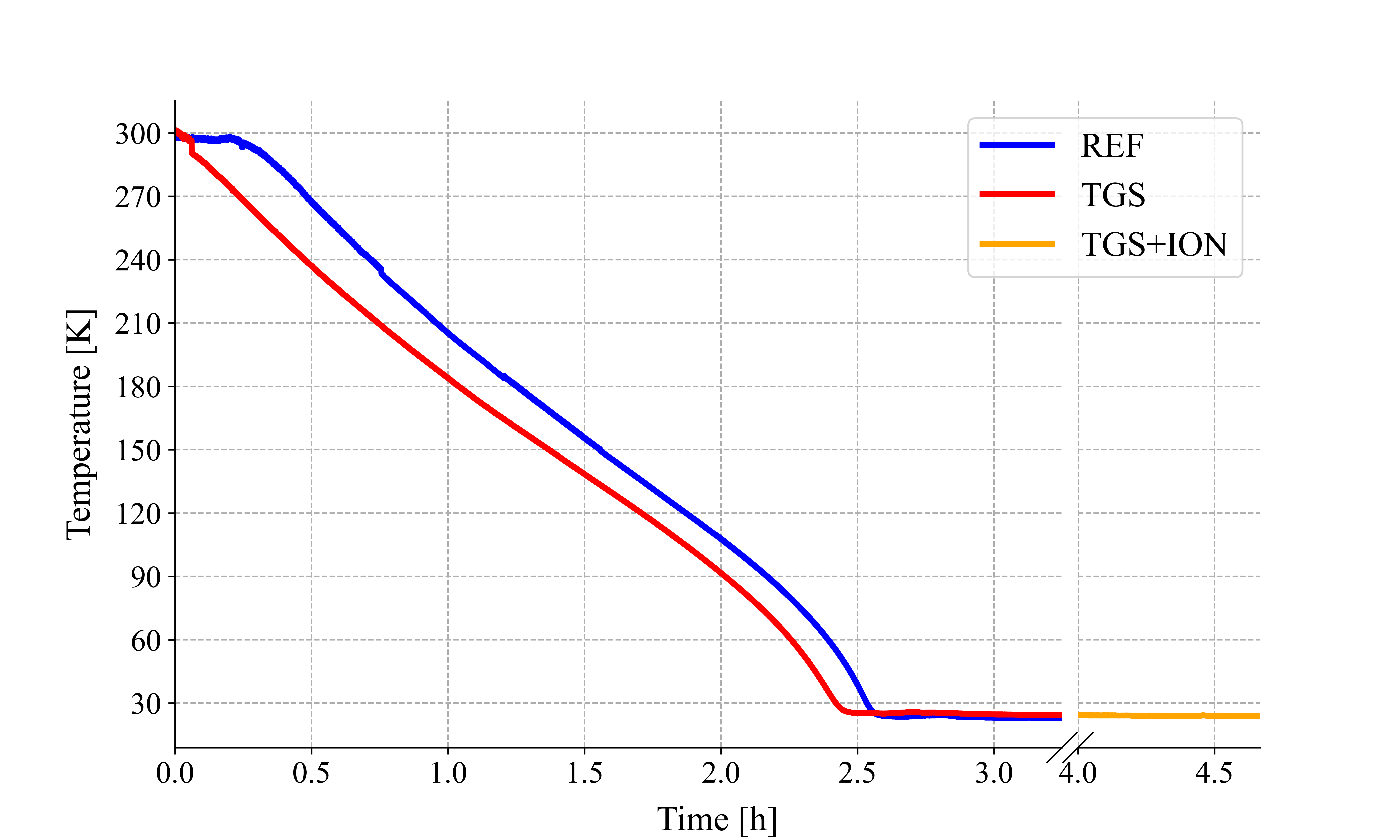}
        \caption{Full}
        \label{fig:cooldown_full}
    \end{subfigure}
    \begin{subfigure}[b]{0.95\linewidth}
        \centering     \includegraphics[width=\textwidth]
        {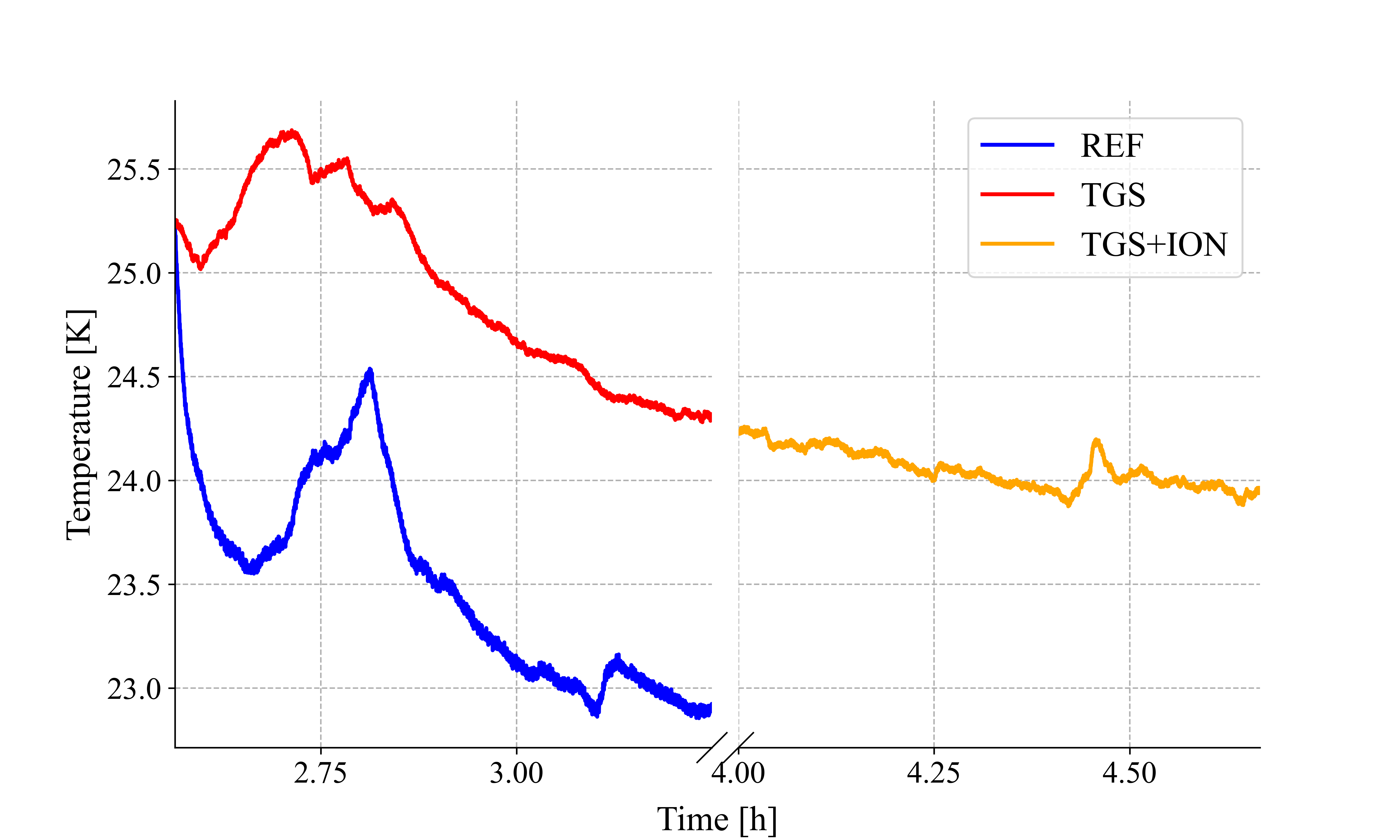}
        \caption{Zoom}
        \label{fig:cooldown_zoom}
    \end{subfigure}
\caption{Cooldown curves of the Cu-110 sample starting at 300\,K. The REF signal includes only conduction, convection, and radiation heat loads.}
\label{fig:cooldown}
\end{figure}

As shown in Figure\,\ref{fig:cooldown_full}, both curve conditions exhibit similar cooldown rates, reaching a base temperature of \( 24 \pm 1 \)\,K. In Figure\,\ref{fig:cooldown_zoom}, the introduction of the ion beam did not significantly perturb the thermal equilibrium, maintaining the temperature around 24\,K with minor fluctuations. These observations suggest that TGS beam heating marginally impacts the cooldown rate, not affecting the final base temperature, and the ion beam interaction remains thermally stable. Notably, the analytical calculation overestimates the sample temperature by approximately 10\,K,. This discrepancy is likely due to the use of steel's emissivity at room temperature, which overestimates the radiative heat transfer at cryogenic temperatures where metal emissivity typically decreases, as well as the assumption of perfect absorptivity for the target holder, further inflating $Q_\text{rd}$ and the overall heat load.

The experiment also recorded the cooldown curve after wrapping the target holder strap in multilayer insulation (MLI) sheath to reduce its absorptivity. This reduced the temperature to 17\,K, confirming the hypothesis that radiative heat is the driving load and increases the base temperature. MLI was not used in further experiments because it restricted the movement of the target holder during alignment for TGS.

\section{Irradiation damage calculation}
\label{appendix:irradiation_damage_calculation}

\subsection{Fluence}

To establish the relationship between irradiation dose and material properties, irradiation time values are converted into corresponding fluence values, which estimate primary radiation damage. Fluence, $\Phi$ ,expressed in $\text{ions}/\text{m}^{2}$, is then a straightforward time integral of the ion beam current, 

\begin{equation} 
\Phi = \frac{1}{q e A} \int I_\text{B}(t) \, dt \label{eq:fluence} 
\end{equation} 

where $I_\text{B}(t)$ is the ion beam current measured in nanoamperes as a function of time, $q$ is the charge of the incident ion, $e$ is the elementary charge, and $A$ is the aperture area in $\text{m}^2$. During the irradiation experiment, the average beam current was 0.45\,nA, making the final fluence $1.9 \times 10^{17}$\,$\text{ions}/\text{m}^{2}$. 
 
\subsection{Displacements per atom}

By applying the result from Eq.\,\ref{eq:fluence} and the average number of vacancies/Å-ion ($K_{\text{SRIM}}$) provided by a SRIM simulation, fluence can be converted into displacements per atom (dpa), 

\begin{equation} 
\text{dpa} = \frac{K_{\text{SRIM}} \times 10^{10}}{N_{\text{Cu}}} \Phi \label{eq:dpa} 
\end{equation} 

where $N_{\text{Cu}} \approx 8.49 \times 10^{28}$\,$\text{atoms}/\text{m}^3$ represents the atomic density of copper, and $K_{\text{SRIM}}$, which is the sum of primary knock-on atoms ($K_{\text{PKA}}$) and recoil atoms ($K_{\text{recoils}}$), is approximately 0.925\,vacancies/Å-ion in the TGS probe depth region. This value is based on the aforementioned SRIM simulation. 

The principal advantage of using dpa as a damage metric lies in its ability to account for the energy of incident ions. This provides a standardized measure of radiation dose, better comparing results across different studies in literature. It should be noted, however, that dose does not equate to damage.
\bibliographystyle{unsrt}

\end{document}